\documentclass[sigconf,authorversion,nonacm]{acmart} 
\usepackage{graphicx} 
\usepackage{xcolor}
\usepackage{hyperref}
\usepackage{amsmath,amsfonts}
\usepackage{subcaption}

\definecolor{OliveGreen}{rgb}{0,0.6,0}
\definecolor{ForestGreen}{RGB}{34,139,34}
\definecolor{myblue}{RGB}{37,165,203}
\definecolor{FAUblue}{rgb}{0.000, 0.2196, 0.3961}
\definecolor{myred}{RGB}{175,32,67}

\newtheorem{theorem}{Theorem}
\newtheorem{observation}[theorem]{Observation}
\colorlet{backgroundcol}{cyan!10!white}

\title{Supercomputers as a Continuous Medium
\subtitle{Deriving the Limits and Laws of Parallel Computing from First Principles}
}
\author{Martin Karp, Niclas Jansson}
\affiliation{
  \institution{KTH Royal Institute of Technology}
  \city{Stockholm}
  \country{Sweden}
}

\author{Philipp Schlatter}
\affiliation{
  \institution{Friedrich-Alexander-Universit\"at (FAU) Erlangen--N\"urnberg}
  \city{Erlangen}
  \country{Germany}
}
\additionalaffiliation{
  \institution{KTH Royal Institute of Technology}
  \city{Stockholm}
  \country{Sweden}
}
\author{Stefano Markidis}
\affiliation{
  \institution{KTH Royal Institute of Technology}
  \city{Stockholm}
  \country{Sweden}
}

\begin{abstract} 
As supercomputers' complexity has grown, the traditional boundaries between processor, memory, network, and accelerators have blurred, making a \emph{homogeneous computer model}, in which the overall computer system is modeled as a continuous medium with homogeneously distributed computational power, memory, and data movement transfer capabilities, an intriguing and powerful abstraction. By applying a homogeneous computer model to algorithms with a given I/O complexity, we recover from first principles, other discrete computer models, such as the roofline model, parallel computing laws, such as Amdahl's and Gustafson's laws, and phenomenological observations, such as super-linear speedup. One of the homogeneous computer model’s distinctive advantages is the capability of directly linking the performance limits of an application to the physical properties of a classical computer system. Applying the homogeneous computer model to supercomputers, such as Frontier, Fugaku, and the Nvidia DGX GH200, shows that applications, such as Conjugate Gradient (CG) and Fast Fourier Transforms (FFT), are rapidly approaching the fundamental classical computational limits, where the performance of even denser systems in terms of compute and memory are fundamentally limited by the speed of light.

\end{abstract}

\begin{document}
\maketitle
\section{Introduction}
Parallel computer models are fundamental tools for computer scientists to assess the potential of computer systems in terms of performance and scalability~\cite{karp1988survey}. Designing a supercomputer and large-scale computing systems requires a careful balance between computing units, memory sub-system, and interconnect to maximize the benefits and match the characteristics of scientific applications, such as linear solvers or Fast Fourier Transforms (FFT)~\cite{ang2014abstract,dongarra2011international,alexander2020exascale}.  By coupling models for algorithmic costs and computer parameters, parallel computer models allow us to estimate and quantify algorithm run times~\cite{williams2009roofline} and align system designs (in terms of compute, memory, and network performance) with the actual algorithmic needs following a co-design methodology. One of the most important advantages of computing models is that they allow us to identify the computational limits of an algorithm running on a computer system in the ideal setting — shedding light on how close we are to the fundamental limits of algorithmic performance.  The fundamental questions computer models can help answer regard: \emph{(i)} the fundamental limitations of classical computer systems in light of algorithmic complexity and \emph{(ii)} how far the performance of a specific algorithm running on a given classical computer system is from these fundamental computational limits~\cite{markov2014limits}. The primary goal of this work is to formulate a computer model that can effectively address these essential questions from first principles, advancing our understanding the computational limits of parallel computer systems.


In a time of rapidly evolving computer systems, characterized by increasing complexity of deep memory hierarchies and heterogeneous technologies, the boundaries delineating computing units, memories, and transport layers are growing less and less distinct. For this reason, conceiving the whole computer system as a continuous entity is becoming an intriguing computer system abstraction. Embracing this notion, we introduce the concept of a continuous, homogeneous computing medium with a well-defined shape and physical extent characterized by a uniform distribution of computational power, memory, where communication within the medium is only constrained by the speed at which messages propagate (speed of light in vacuum). We denote this continuous computing medium as the \emph{homogeneous computer model}. This approach allows us to assess a computer's performance based on critical factors such as computing density, bandwidth to an infinite external memory, local memory density, and the shape and volume of the computer system. 

In this work, we focus on the performance modeling of conventional algorithms when executed on an abstract, spatially continuous computer system. We evaluate the model in the conventional digital setting where the computer system can efficiently amortize the cost associated with scientific algorithms, encompassing data movement, floating-point operations, and intra-computer communication. Using the homogeneous computer model, we can explain how well-established principles in parallel computing, including Amdahl's and Gustafson's laws~\cite{amdahl1967validity,gustafson1988reevaluating} can be generalized. By incorporating costs associated with the I/O complexity of parallel algorithms~\cite{jia1981complexity}, the homogeneous computer model can predict the behavior in the ideal case, and determine the performance limits, constrained by classical physical limits. The computational cost of algorithms such as the Fast Fourier Transform (FFT), matrix multiplication, and iterative solvers has been extensively studied~\cite{ballard2014communication} and is well understood. By using I/O complexity analysis and the homogeneous computer model, the limit of performance and scaling behavior for common algorithms can be evaluated, and other computer models, such as the roofline model, and parallel laws, such as Amdahl's and Gustafson's laws, can be recovered.

Our objective is to assess the performance limits of scientific algorithms and computer systems from first principles, relaying the homogeneous computer model. We make the following contributions:
\begin{itemize}
    \item We introduce a continuous, homogeneous computer model that generalizes discrete parallel computer systems in the limit of a very large number of interconnected computing devices. We then apply this model to a range of scientific algorithms, showing their performance characteristics as we change various computer parameters.
    \item Within the homogeneous computer model, we show how parallel scaling laws, including Amdahl's and Gustafson's, are formulated.
    \item Finally, we model the performance of the Frontier, Fugaku, and the Nvidia DGX GH200 supercomputers as well as a supercomputer composed of a single large die based on the Nvidia A100 and evaluate it compared to a homogeneous computer with a $10^9$ larger computational density.
\end{itemize}


\section{Related Computer Models}\label{sec:models}
In this section, we cover the main aspects of related work to the homogeneous computer, both with regard to the abstract view of a computing device, and models that are used to analyze and develop algorithms in the field.

\noindent \textbf{I/O complexity models.} To evaluate the cost associated with moving data across the memory hierarchy, models such as the parallel random access machine (PRAM)~\cite{karp1988survey}, where multiple processors access one global external memory in parallel, have been used. The parallel external memory model (PEM)~\cite{arge2008fundamental} extends this notion and also considers that each processor has a small private memory. Similarly, the so-called I/O cost, $Q$, (in terms of loads/stores to external memory) of algorithms has been analyzed with the Red-Blue pebble game as proposed by Yie and Kung~\cite{jia1981complexity,demaine2018red} and it has been used to obtain tight bounds on the I/O cost for several numerical algorithms~\cite{kwasniewski2019red,smith2017tight}. The homogeneous computer is similar to PEM in the limit of an infinite number of processors and we use the costs from previous works to provide insight into how the I/O-cost $Q$ can be amortized within a homogeneous computer and model the time needed for data movement. 


\noindent \textbf{Horizontal communication models.} 
For communication between processors (horizontal communication), models such as the postal model and the family of LogP models to model the communication time have been proposed~\cite{rico2019survey,culler1993logp,hoefler2010loggopsim}. Parameters such as congestion, noise, and message size have been incorporated into these. In this work, we focus on the best-case propagation time of a message through the computer, capturing the fundamental communication limit. This aligns with early works about horizontal communication where it was observed that the propagation speed becomes the limiting factor in the ideal case~\cite{fisher1988your}.  

\noindent \textbf{The roofline model.} Computer models are also used to estimate the run time of a parallel computation as the complexity of the processor and software make simplifications necessary. One commonly used model is the roofline model~\cite{williams2009roofline}, where the computational speed of the parallel computer in terms of bandwidth and floating point performance is used to obtain a bound on the fastest run time of a program according to 
\begin{equation}
    T = \max \left (\frac{W}{\pi}, \frac{Q}{\beta}\right)
\end{equation}
where not only the I/O cost of the algorithm is considered, but also the work in terms of the number of operations necessary $W$ and the performance in operations per time unit $\pi$ together with the bandwidth $\beta$ available to move the I/O cost. Related to this model is the operational/computational intensity of a program, described by $I=W/Q$. The roofline model has been extended in several regards, and in particular, this work is similar to models that extend the roofline model to also take into account horizontal communication costs~\cite{dufek2021extended,cardwell2019extended}.

All of these models have been used to drive algorithmic developments and relate the cost of an algorithm to the execution time. In the numerical linear algebra domain, the study of communication-avoiding algorithms (focusing on minimizing the communication between processors and across the memory hierarchy) has developed into a major scientific field as communication is the primary limiting factor on current and future computer systems~\cite{ballard2014communication,abdelfattah2021survey,dongarra2011international}. Often the models where the execution time is composed of the time to move data from memory, execute floating point operations, and communicate between processes according to $T=T_W+T_Q+T_L$ have been used \cite{solomonik2017trade,hoemmen2010communication}. In our work, we follow the same line of thought, but apply it to a continuous, homogeneous computer. 

\noindent \textbf{Continuous computer models.} 
All of the mentioned approaches consider discrete messages and processors. The motivation for this work is to consider the case where computational devices grow smaller and computation is getting dispersed throughout the computer. In the limit of very many processors, an alternative approach is, therefore, to consider the computer as continuous, and spatial models of computation might provide more insight. Two models in particular are of interest in our work, namely amorphous computing, where the computer can be considered a continuous mass of microscopic computers, and computation fields which also consider the spatial extent of the computer to be continuous~\cite{abelson2000amorphous}. Several works have brought forward this view of computing, but in particular, focused on how to coordinate and program such systems. Our work instead focuses on what the fundamental performance limits and behavior of conventional algorithms would be on such a system~\cite{nagpal2004engineering,beal2004programming}. Other works from the perspective of physical limits of computing can be illustrated by the \emph{Ultimate Laptop} as proposed by Lloyd~\cite{lloyd2000ultimate}, where the fundamental computational power of a 1kg mass was considered, and the computation was uniformly dispersed in the entity. While the computational limits of Lloyd are nowhere near being broken~\cite{markov2014limits}, the idea of the ultimate computing machine being completely homogeneous is partially what led to the model under consideration in this paper, albeit limiting ourselves to classical physics and avoiding effects due to quantum mechanics. In line with this thought, we use our computer model to evaluate the best possible performance achievable for a digital computer with a certain computing density.

This work differs in that we incorporate aspects of all the above models and relate them to the run time/performance of considered algorithms in a continuous computer setting. Through the homogeneous computer model, it is possible to relate parameters to algorithmic costs and the performance and scaling behavior of computational workloads as we approach an even larger scale and the fundamental limits of computation~\cite{markov2014limits}.
\begin{figure}
    \centering
    \includegraphics[trim={3cm 3.18cm 2.5cm 2.5cm},clip,width=0.56\columnwidth]{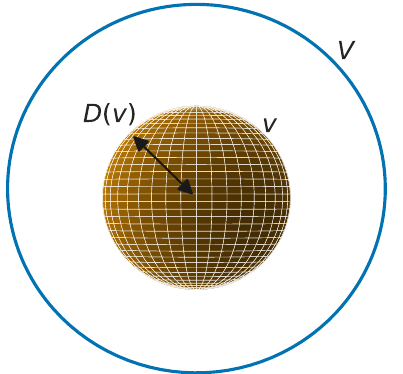}
    \caption{A homogeneous computer with an active volume $v$ with distance function $D(v)$ within a spherical domain $V$. The volume has a density of $\pi$, $\beta$, and $s$ for the computational power, bandwidth, and local memory size respectively.}
    \label{fig:hom_computer}
\end{figure}
\section{The Homogeneous Computer Model}\label{sec:hom_comp_model}
The defining feature of the homogeneous computer model, which sets it apart from other parallel models like PEM, is its departure from discrete computer system components in favor of a continuous computational volume denoted as $V$, within which algorithms operate. The volume is then of a certain geometry, dictating the distance messages need to travel within some sub-volume $v \leq V$, and we illustrate this idea in Figure~\ref{fig:hom_computer} for a sphere. In this work, we consider that each sub-volume follows the same distance metric $D(v)$. Using this model we then go on to evaluate how different algorithms, often used in large-scale parallel computing could perform and predict how future computer systems, with the computational power spread out throughout the system might behave. Depending on the selection of the parameters in the model we aim to capture the best possible scenario for a given computer and give predictions of the performance and execution time in an ideal case.

By homogeneous, we mean that the parameters for performance and the bandwidth are uniformly distributed throughout the domain and are linear functions of the computing volume, and the only non-linearity introduced is through the shape of the computer volume and depending on the algorithmic cost. For the distance function of the volume, the only constraint is that it is monotonically increasing with $v$. The model has the following parameters:
\begin{itemize}
    \item The total volume $V$ and a corresponding distance function $D$ determined by its shape. If only a subset of the volume, $v$, is used, the distance of the point furthest away from the center of the active volume is parameterized by a function $D(v)$.
    \item The entire volume has a performance $\Pi$ and a compute density $\pi$ per unit volume.
    \item A local memory $S$ that is uniformly distributed in the volume with a memory density $s$ per unit volume.
    \item Information propagates at a speed $c$ within the volume.
    \item The volume is uniformly connected to an external infinite memory with a total bandwidth $B$ and a bandwidth of $\beta$ per unit volume.
\end{itemize}

Executing an algorithm within the homogeneous computer framework is linked to the algorithm's cost, which directly corresponds to the computer parameters described earlier. This cost is characterized by the following critical parameters:

\begin{itemize}
    \item $Q$: I/O cost, denoting the amount of data needed to be moved from the infinite global memory.
    \item $W$: Computational work required to execute a given algorithm. 
    \item $L$: Latency, or communication cost, the wavefront for any output parameter, i.e. how large part of the input volume any output variable is dependent on.
\end{itemize}

The I/O cost, we consider in this work, is similar to the I/O cost considered in the Red-Blue Pebble Game~\cite{jia1981complexity}. In this paper, the considered computer is a classical digital device, and as such the cost is associated with the number of bits needed per word. At the start of each program, the entire problem is assumed to be uniformly dispersed in the infinite external memory and local memory.

Of note, is that the costs here also depend on the computer parameters, in particular, the I/O cost depends on the size of the small fast memory $sv$ and the latency $L$ on the volume used for the computation. To be precise we have: $Q=Q(sv)$, $L=L(v)$.

This leads to the model having a more complex relationship between the computing volume and the capabilities of the computer. The homogeneous computer can thus capture complex relations between the algorithm and hardware through the computational cost, even though the computer itself is homogeneous. Considering the run time $T(v)=f(v)$ we now have that for an algorithm being executed on a volume $v$ the run time is
\begin{equation}\label{eq:def}
\begin{split}
     f(v) &= T_W(v) + T_Q(v) + T_L(v) \\
     &T_W(v) = \frac{Q(sv)}{\beta v} \\
     &T_Q(v) = \frac{W}{\pi v}\\
     &T_L(v) = \frac{D(L(v))}{c}.
\end{split}
\end{equation}
While the different parts of the algorithm might be able to overlap, similar to the roofline model, the approximate run time is our primary interest in this work. In particular, we consider the minimization problem of the run time (or some other metric $f(v)$) for a given computer and algorithm
\begin{equation}\label{eq:modeltime}
\begin{split} &\underset{v}{\text{minimize }}   f(v) = \frac{Q(sv)}{\beta v} +\frac{W}{\pi v}+\frac{D(L(v))}{c} \\
   &\text{such that } v \le V, 
\end{split}
\end{equation}
letting $f(v)$ have continuous derivatives simplifies the optimization compared to solving for $f(v) = \max \left \{ \frac{Q(sv)}{\beta v} ,\frac{W}{\pi v},\frac{D(L(v))}{c} \right \}.$ The optimization problem will be important to properly capture the best-case performance as in some cases using a smaller portion of the volume yields a better solution than using the entire computer volume. A prime example of this arises when the time needed to send messages greatly exceeds the time to execute the computation, making it more advantageous to work with only a fraction of the available computational space.

\noindent\textbf{Units in the Homogeneous Computer Model.} No specific units are imposed on the parameters and cost within the homogeneous computer model and they can be modified to accommodate various computers and algorithms. To illustrate the use of the model, we focus on conventional scientific algorithms in floating point precision and consider flop/s, words/s, and m/s as our units for $\Pi, B$, and $c$.

Regarding the volume of the computer, we start by considering the 3D case, where the diameter and thus the distance function $D$ scales as the cube root. Extending the model to more dimensions and geometries can be done by modifying the distance function $D$ as a function of the volume.

In addition, the units in the computer model can be chosen for any suitable metric of cost and function to minimize. To start we focus on the run time, which also gives us the performance in flop/s as $W/f(v)$, but we also envision that the cost can be connected to other metrics $[f(v)]$ such as the energy consumption or monetary cost of the computation.

\noindent\textbf{Considerations on Non-von Neumann Architectures and Latency Cost.} Non-von Neumann computer architectures might at first glance not be accommodated within the homogeneous computer, but in practice, they can often be regarded as processors with large local memory, $S$, and perhaps a different performance metric than flop/s. They can therefore also be assessed within the homogeneous model. While we consider floating-point arithmetic in this work, there is no limitation on considering other forms of computation or introducing another metric for the costs.

Another consideration is the communication or latency cost. As it stands, the communication is, in comparison to conventional models for communication, rather optimistic for the distances and volumes we consider in this paper. However, to capture the ideal performance behaviors we focus on the communication speed. We foresee that finding more realistic distance metrics as a function of $v$ for the cost will make it possible to more accurately capture the latency time and also capture several conventional communication models such as LogP and similar~\cite{rico2019survey}. 

A different aspect that is neglected in the model is the issue of communication patterns. Similar to the latency cost a more exact definition of the distance metric and latency cost would in a way accommodate this, but in the current state the model provides an optimistic bound where the computation is not limited by more complex non-aligned accesses. Overall, the proposed model is best used as a way to illustrate the fundamental limits of performance and reason around the intrinsic relation between computational cost, scaling, and computer parameters.

\section{Methodology}\label{sec:modeling}
We focus on three conventional scientific computations and evaluate their performance limits within the homogeneous computer model. The choice of applications correlates with tests used for the Top500 such as HPL and HPCG, albeit with slightly lower computational costs, and aims to capture the computational limits, both as size and performance increases. The three scientific computational kernels are:

\begin{itemize}
\item \textbf{Matrix Multiplication.} We consider conventional matrix multiplication (MxM) of two square matrices $A B = C$ with dimensions $n \times n$. The bounds with regards to the vertical communication cost $Q$, have been of extensive study and several bounds have been proposed. We use the bounds from \cite{smith2017tight} for non-Strassen-like algorithms for matrix multiplication. For the least amount of communication in the neighborhood we consider the optimal case where $C$ is uniformly distributed in the computer and only needs to obtain data from $\frac{1}{n}$ part of the computer volume.

\item\textbf{The Conjugate Gradient Method.} The conjugate gradient method (CG) is used to solve, the positive semi-definite, symmetric linear system of the form $Ax=b$. As has been shown several times, the I/O cost of CG does not decrease more than linearly with the size of a small fast memory $S$~\cite{ballard2014communication}, this can be seen in Table \ref{tab:costs} where we list all the costs for our considered algorithms and we consider one iteration of CG. To provide an optimistic bound on the execution time of CG we omit the costs associated with the evaluation of $A$, as this can vary significantly between linear systems. The cost $L$ is therefore only caused by the global reductions at each iteration of the algorithm. The costs $W, Q$ are taken from~\cite{karp2022reducing}.

\item\textbf{Fast Fourier Transform.} We consider the 1D FFT with $W$ and $Q$ obtained from \cite{takahashi2019fast}. The horizontal communication that occurs in FFT is $v$ as there is a dependence from each input to each output in the computational graph. 
\end{itemize}

For our experiments, we solve the minimization problem of \eqref{eq:modeltime} for the algorithms above. Each of these algorithms has been studied to compute lower bounds on data movement, but in the case of matrix multiplication, also to decrease flops with Strassen-like approaches.  To investigate the bounds of communication of these algorithms we use some best-case computational costs for a given computer and algorithm. The bounds are optimistic and it should be noted that the volume and communication speed reduce to each other (a less dense computer with a slower communication speed is in the homogeneous computer is closely related to a denser computer with a faster communication speed). Therefore, we use the speed of light $c \approx 3\cdot 10^8$ m/s for our calculations to estimate the limit for a computer of a certain volume and an associated distance metric, all absolute values in our discussion are in this way normalized by the speed of light and should be viewed as optimistic performance bounds on any given algorithm. 

\begin{table}[t]
\caption{Costs of our considered algorithms as functions of problem size $n$.}\label{tab:costs}
\begin{center}
\begin{tabular}{llll}
\toprule
Algorithm &  $Q(n,S)$ & $W(n)$ & $L(v,n)$\\\midrule
MxM & $\frac{2n^3}{\sqrt{S}} -3S $& $2n^3$ & $\frac{v}{n}$\\
FFT& $\frac{2n\log(n)}{\log(S)}-2S $& $\frac{8}{3} n \log(n) $&$v$\\
CG & $7n-4S$ & $ 17n$&$ 2v$\\
\bottomrule
\end{tabular}
\end{center}
\end{table}
\begin{table}[t]
\caption{Range of computer parameters and problem sizes.}\label{tab:params}
\begin{center}
\begin{tabular}{ll}
\toprule
Parameter  &  Range \\\midrule
$\pi$ & $[10^{-30},10^{30}]$ \\
$\beta$ &$[10^{-30},10^{30}]$ \\
$c$ & $3\cdot10^{8}$ \\
$s$ & $[10^{-30},10^{30}]$\\
$D(v)$ & $v^{1/3}$\\
$V$ & $[10^{-14},10^{14}]$\\
$n$ & $[1000,10^{30}]$\\\bottomrule
\end{tabular}
\end{center}
\end{table}
The minimization problem in \eqref{eq:modeltime} is solved using Python and the optimization module in \texttt{SciPy}, with the bounded Brent method, ensuring the constraint $0< v \le V$. The code is released open-source as a Jupyter notebook\footnote{URL shared upon publication}.

To study the homogeneous computer we consider several different compute densities, bandwidths, problem sizes, and volumes and we summarize them in Table~\ref{tab:params}. For the units we focus on flop/s, word/s, and the distance is in m.  We used between 10-20 different values across several orders of magnitude for the different parameters. The optimization problem was solved for the entire range in a couple of minutes on a personal computer for each algorithm. Looking at the predicted results over a wide range of parameters we seek to capture the performance limits of any given computer across a large range of scales.

\section{Results}
\label{results}
\begin{figure*}
    \centering
    \includegraphics[trim={0 0.5cm 0 0},clip,width=2\columnwidth]{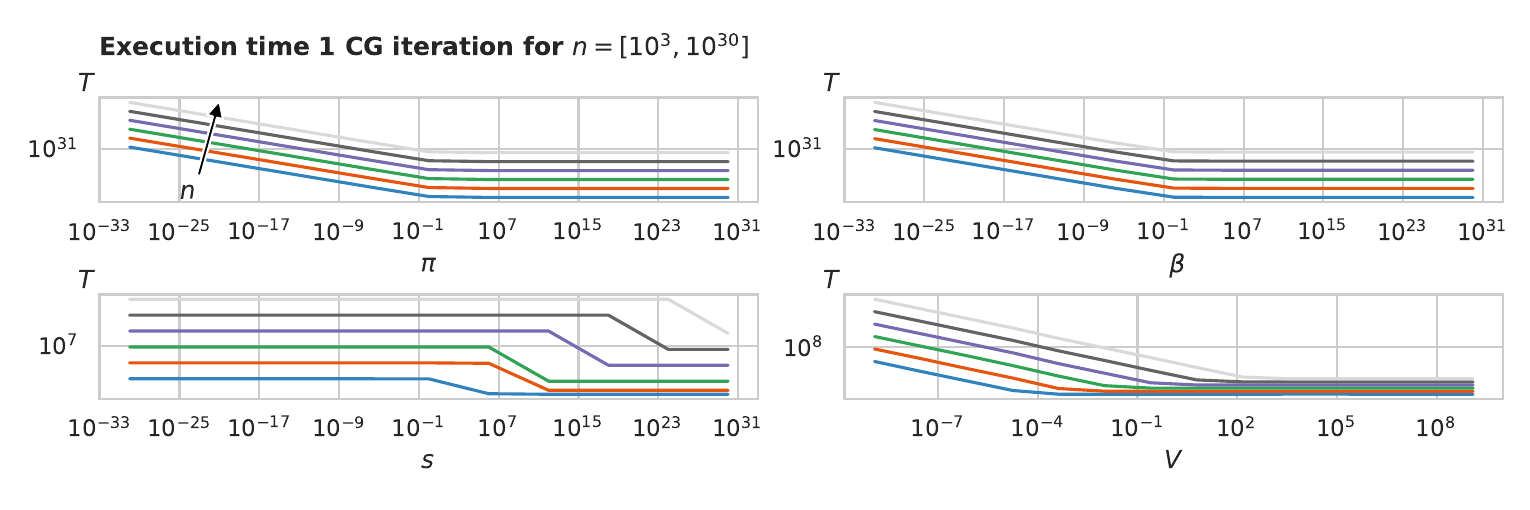}
    \caption{The run time $T$ varies depending on the compute density $\pi$, bandwidth density $ \beta$, memory density $s$, and the computer volume $V$. Each line corresponds with one value of $n$, the problem size.}
    \label{fig:1d_plots}
\end{figure*}

\subsection{Run Time}
In our first assessment of the behavior of the homogeneous computer, we show the run time of CG for various problem sizes $n$ in Figure~\ref{fig:1d_plots}. These illustrate clearly how the run time of the algorithm is affected by an increase in $\pi,\beta,s$, and volume $V$, all other parameters held equal, until a certain point when they are no longer the limiting resource. At this point, the performance does not increase regardless if the performance of the computer is increased for this parameter. Throughout this discussion, we will define performance as $W/T(v)$. However, as noted earlier, there is a coupling between all of the parameters and the computational cost, it, therefore, makes sense to also view the performance of the computer in several dimensions, in particular how the run time changes with $\pi, s$ for a fixed $\beta$. 

\begin{figure*}
    \centering
    
    \subfloat[MxM\label{1a}]{%
       \includegraphics[width=0.333\textwidth]{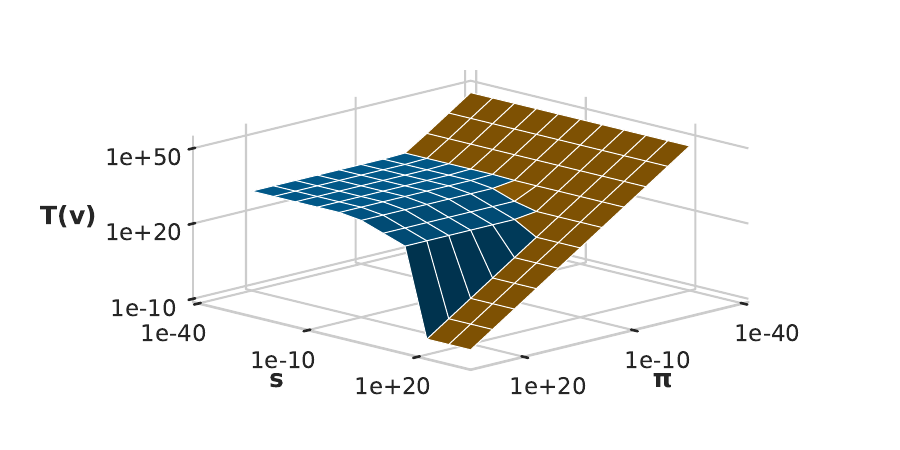}}
  \subfloat[CG\label{1b}]{%
        \includegraphics[width=0.333\textwidth]{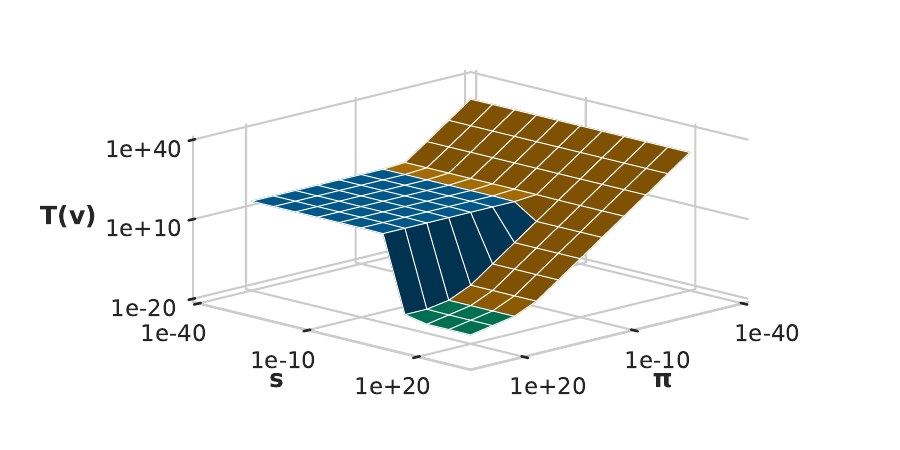}}
  \subfloat[FFT\label{1c}]{%
        \includegraphics[width=0.333\textwidth]{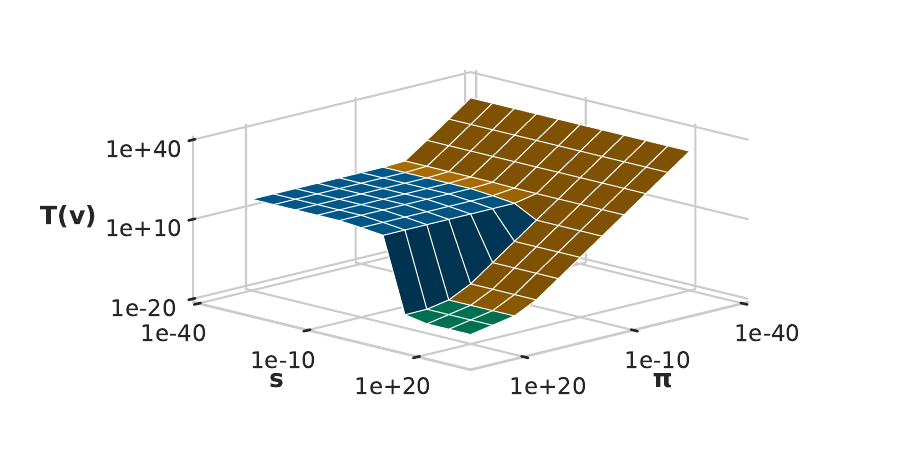}}
    \caption{Run time of CG, MxM, and FFT as the memory density $s$ and performance $\Pi$ increase. The performance is impacted primarily by the latency in the lowest corner and as the performance density increases the active volume $v$ decreases and so does the run time.  For a smaller problem, increasing the performance is more important, as the effect of caches is much larger and we are not as limited by the bandwidth. A blue surface corresponds to $T_W$ taking the most time, yellow-brown for $T_Q$, and green for $T_L$.}
    \label{fig:3d_perf}
\end{figure*}

As $s$ and $\pi$ of the homogeneous computer increases, the performance increases and run time decreases, until the computer is limited by $\beta$ or $c$. This is evident in Figure~\ref{fig:3d_perf} where the execution time of the different algorithms is shown for a range of computer parameters. As can be seen, the range of values for the computation is tens of orders of magnitude and the figures are not meant to find exact values, but illustrate the behavior of the computer and the different algorithms as the different parameters change. We have also colored the sections of the figures where $T_W$ is the limiting factor in yellow-brown, the memory time $T_Q$ is the limiting factor in blue, and the latency $T_L$ in green. Using common vocabulary, we color whether the computation is compute-bound, memory-bound, or latency-bound in different colors. In some sense, the performance plots shown here can be thought of as the roofline model in three dimensions, where we have colored the different regions depending on the main computational bottleneck. If we let the volume be constant, the model reduces and replicates the same plots one would get from the roofline model, albeit with additive contributions from $T_w, T_Q$ as we do not consider overlap.

Overall, we also see the different behaviors of the different algorithms under consideration, something that is also mirrored in the Top500, comparing for example how HPL and HPCG behave as system size and performance increase~\cite{dongarra1997top500}. The performance of CG method and FFT is notably similar, compared to dense MxM, and is much sooner limited by memory bandwidth and latency. This similarity is due to that, while $W,Q$ differs for the two algorithms, $I=W/Q$ omiting logarithmic factors is notably similar as well. In particular, the absolute performance achievable for CG and FFT has an almost negligible effect on both bandwidth and performance, as the cost of the global reductions is comparably high to $I$. Even as we are operating with a communication speed equaling the speed of light, the impact of communication within the volume for CG and FFT quickly becomes a computational bottleneck as for most problem sizes and computer volumes This effect is especially amplified in the limit of high $\beta, \pi$, and very many processors.
\begin{observation}
The model provides the best-case scenario for a chosen metric $f(v)$, based only on first-principles observations regarding the computing density and shape of the computer. It provides a visual way of reasoning around the best-case performance in multiple dimensions depending on $\pi, \beta, s$ and how the performance transition to be limited by $T_L, T_W,$ or $T_Q$. The model can steer the development and assess fundamental performance limitations to new algorithms and computer systems.
\end{observation}

\begin{figure}
    \centering
    \includegraphics[width=\columnwidth]{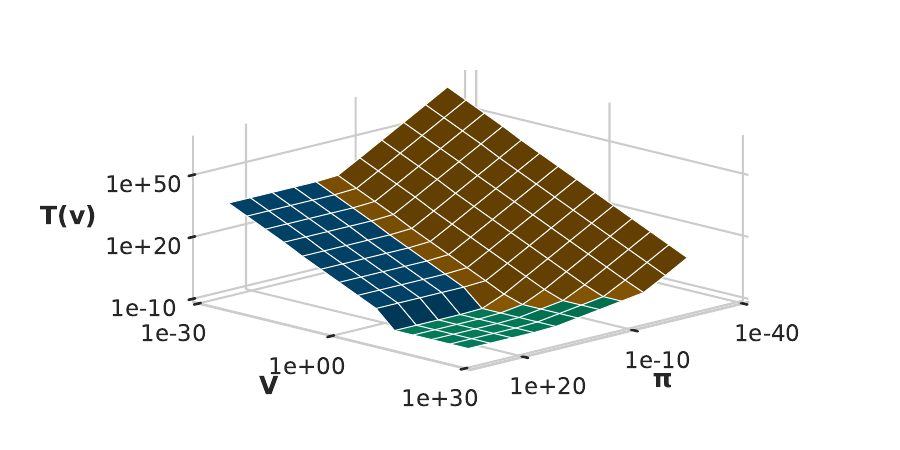}
    \caption{Run time for CG as the volume of the computer $V$ and performance $\Pi$ increase. The performance is impacted primarily by the latency in the lower corner and as the performance increases the active volume $v$ decreases and so does the run time.  For a smaller problem, increasing the performance is more important, as the effect of caches is much larger and we are not as limited by the bandwidth.}
    \label{fig:3d_scaling_perf}
\end{figure}
\subsection{Parallel Scaling \& Efficiency}
Scaling, in the context of a homogeneous computer model, involves maintaining the same computing medium while expanding its volume, akin to augmenting the number of processors in traditional computing setups.  We show the performance for CG when increasing the computer volume in Figure~\ref{fig:3d_scaling_perf} where the execution time decreases until using a smaller part of the volume gives the best trade-off between latency, bandwidth, and floating point operations. Likewise, we can use the model to evaluate the parallel efficiency of our approach by expanding in volume while maintaining constant compute density. We start with a strong scaling analysis, where we maintain uniformity in all the medium properties, including computer parameters and algorithm settings, while solely increasing the volume.

\begin{figure}
    \centering
    \includegraphics[width=\columnwidth]{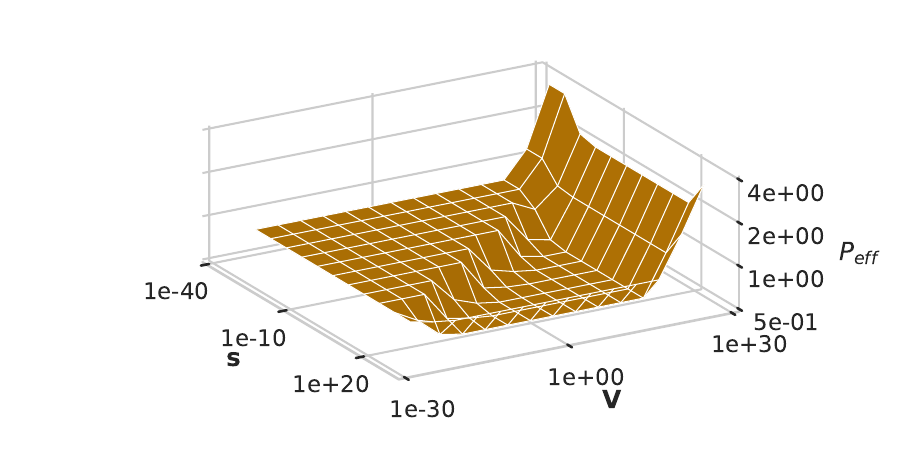}
    \caption{An increased volume (and different memory densities $S$) changes the parallel efficiency when $\pi \approx \beta$ for FFT.}
    \label{fig:fft_par_eff_work}
\end{figure}

\begin{figure}
    \centering
    \includegraphics[width=\columnwidth]{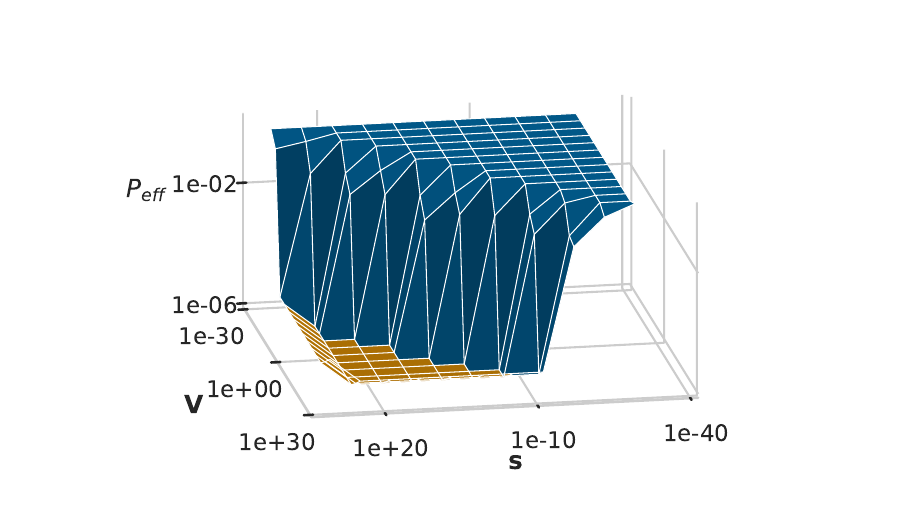}
    \caption{An increased volume (and different memory densities $S$) changes the parallel efficiency when $\pi \gg \beta$ for FFT.}
    \label{fig:fft_par_eff_mem}
\end{figure}

We can define the parallel efficiency in our case, for our metric $f(v)$, as
\begin{equation}\label{eq:par_eff}
    P_{\text{eff}} = \frac{f(v_0)v_0}{f(v)v},
\end{equation}
where we have introduced the starting volume for the scaling $v_0$, which can be thought of as the sequential baseline. We must stress though that the homogeneous computer is parallel, and we let a computation using some volume $v_0$ be the baseline performance, and for scaling it holds that $v\ge v_0$. Of interest is that the computational cost of the algorithm with regards to $W$ is constant, while the I/O cost greatly decreases with an increase in $sv$, and when the computation is the limiting factor, such as what is shown in Figure~\ref{fig:fft_par_eff_work} a parallel efficiency above 1 (super-linear scaling) can be achieved compared to the smallest volume used in the computation for FFT. The same is true when the memory is the limiting factor as shown in Figure~\ref{fig:fft_par_eff_mem}, where we see a similar trend, but less amplified. This is due to that when the small memory becomes sufficiently large the performance of the algorithm does not improve by adding more volume. This might seem to be a contradiction to Amdahl's law, but as we noted it is because $Q(s)$ is dependent on the small memory. As such the computational cost of the algorithm actually decreases with the volume unlike the case in Amdahl's law where the parallel and sequential portions of the program are held constant as the number of processes increases.
Instead, investigating the weak scaling, where the computational cost increases with the number of processors, it can be formalized as
\begin{equation}\label{eq:par_eff}
    P_{\text{weak}} = \frac{f(v,n)}{f(v_0,n_0)}
\end{equation}
within the homogeneous computer model. For weak scaling, we keep some metric depending on the problem size $K(n)$ constant per unit volume s.t. $K(n)/v=K(n_0)/v_0$. How $K(n)$ is chosen depends on the relevant weak scaling parameter for the specific problem, and choosing which $K$ to hold constant is an issue in the weak scaling setting. One observation is that $K(n)$ can be chosen for any algorithm to indicate good scalability, while it might not be connected to any relevant scientific output. 
\begin{observation}
As the I/O cost $Q$ is a decreasing function of $sv$, strong scaling to a larger volume $v$ can yield super-linear speedup as $\beta,\pi$ can be utilized more efficiently. For weak scaling, choosing what parameter $K(n)/v=K(n_0)/v_0$ to hold constant has a large effect on the scaled performance. Choosing which $K(n)$ to hold constant when executing benchmarks is not an objective decision. We suggest a standardization would be to keep the output size per unit volume constant.
\end{observation}

\subsection{Deriving Amdahl's and Gustafson's Laws from First Principles}
These first results shown in Figures~\ref{fig:fft_par_eff_work} and  \ref{fig:fft_par_eff_mem} with regard to strong and weak scaling give us a hint about formalizing Amdahl's and Gustafson's laws in the context of the homogeneous computer model. First, by considering the problem size fixed and independent of the number of processors we have Amdahl's law, often written as
\begin{equation}
    \text{Speedup} = \frac{1}{t + \frac{p}{N}} = \frac{1}{\frac{1}{N} +(1-\frac{1}{N})t}\quad \text{for constant } t,p
\end{equation}
which bounds the possible speedup for a program with a sequential portion $t$ and a parallel portion $p = 1-t$ when executed on $N$ processors. Speedup for a constant input would in the homogeneous computer model be generalized as the following
\begin{equation}
    \text{Speedup} = \frac{1}{\frac{v_0}{v} + (1-\frac{v_0}{v})\frac{T_L(v_0)}{T(v_0)} }
\end{equation}
and Amdahl's law would be the special case where the parallel fraction (and thus also the sequential fraction) 
\begin{equation}
    p=\frac{\frac{Q(sv)}{v\beta}+\frac{W}{v\pi}}{T(v)}
\end{equation}
is constant, from this we have that the highest possible speedup given by
\begin{equation}
\text{Speedup} = \frac{T(v_0)}{T_L(v_0)} \quad \text{as } v_0 \le v.
\end{equation}
A more accurate model of the achievable speedup is $T(v_0)/T_L(v)$ as $T_L(v)$ is monotonically increasing with $v$. With this in mind, it is clear the homogeneous computer model proposed captures Amdahl, but also gives insight into that the cost associated with the parallel portion can decrease as $v$ increases and that the case when the parallel portion is held constant is only a special case. In particular, this constraint on $t,p$ as the computing volume increases, leaves Amdahl's law pessimistic in its predictions of achievable speedup, also for the homogeneous computer. The best-case speedup regardless if $p$ is constant remains the same in all cases, posing a limit on how far we can strong-scale.

That Amdahl's law was pessimistic about performance improvements possible with parallel computation was originally noted by Gustafson in 1988~\cite{gustafson1988reevaluating} as most problems change and the problem size and computational cost increases with the number of processors. He therefore proposed the scaled speedup is of greater importance than the absolute decrease in execution time. Gustafson's law instead considers that the problem is scaled with the number of processors as in weak scaling and is often written as

\begin{equation}
    \text{Scaled~speedup} = \frac{t + pN}{t+p} = t + pN = N + (1-N)t
\end{equation}
where $K(N)/N$ is held constant as the number of processors $N$ increase. Gustafson's law in the context of the homogeneous computer model we could formalize instead as the following, where the speedup is described as
\begin{equation}
\begin{split}
    \text{Scaled-speedup} &= \frac{T_L(v_0)}{T(v_0)} +p \frac{v}{v_0}\\ 
    &= \frac{v}{v_0} + \left (1- \frac{v}{v_0}\right )\frac{T_L(v_0)}{T(v_0)}
\end{split}
\end{equation}
with the smallest sequential time being~$t=T_L(v_0)/T(v_0)$, and the computational problem size scales as $K(n)$. Of note is that when weak scaling within the homogeneous computer, in for example a spherical volume, the performance increases with the volume, and the diameter increases at a slower rate, meaning that when a larger part of the volume is used, the distance needed to be traveled increases at a slower than the computational capabilities of the computer. If $K(n)=n$ is held constant the behavior changes drastically from when for example $K(n)=Q(n)$ is kept constant per unit volume. As algorithms such as matrix multiplication scale as $n^3$, keeping $n/v$ constant will then yield significantly higher scaled speedup than keeping $K(n)=n^3$ constant, illustrating that a rather simple concept such as weak scaling is inherently subjective. 

Overall, with regard to scaling, we see that the problem size, algorithmic cost, and computer parameters all interact as we increase the number of parallel computations, making fair scaling comparisons challenging, especially for weak scaling. This is evident in the larger field of parallel computing where defining a fair baseline is subject to differing opinions about which $K(n)$ to hold constant. Based on our results in the paper, across different algorithms, a suggestion for the least subjective choice for $K(v)/v$ is to keep the size of the output per unit volume constant rather than the computational cost. The homogeneous computer model shows how these non-trivial interactions are due to the nature of how the computational cost, computer parameters, and communication requirements all impact each other.
\begin{observation}
    In the homogeneous computer model, the sequential portion of the program depends on $T_L$, and Amdahl's law corresponds to the special case where the parallel and sequential portions of the program are held constant reduces to the special case when $\frac{T_Q+T_W}{T}$ does not change as a function of $v$. Gustafson's law translates in a similar way, but is in the best case also limited by the smallest sequential portion $T_L(v_0)/T(v_0)$ for a chosen $K$.
\end{observation}


\subsection{Homogenized Supercomputers and Computational Limits}\label{sec:mod_comp}
In this subsection, we map the homogeneous computer model to current technology and extract the corresponding parameters of the model to those found in current systems. Evaluating for example the Nvidia A100, built with a 7nm process and a die size of 826mm${}^2$~\cite{ampere} we have a peak performance in double precision of close to $30$ Tflop/s, a bandwidth of $1550$ GB/s, and an L1+L2 cache size of $60$ MB. This gives $\pi =3.6\cdot 10^{16}$ flop/(m${}^2$s), $\beta = 2.3\cdot 10^{14}$ word/(m${}^2$s), and $s=9.2\cdot10^9$ word/m${}^2$ with one word being an IEEE double precision number, for a square homogeneous computer composed of a medium with the same parameters as the Nvidia A100. 

\begin{figure}[t]
    \centering
    \includegraphics[width=\columnwidth]{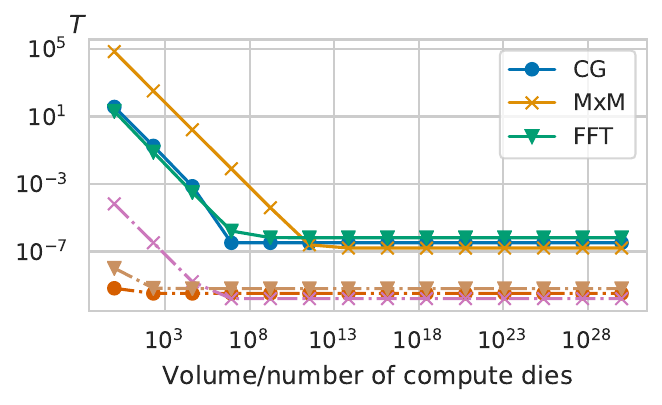}
    \caption{Performance for a square homogeneous computer with the same parameters as the compute die of the Nvidia A100. We keep the number of outputs constant across the different algorithms, using $n=10^{12}$ for FFT and CG and $n=10^6$ for MxM. We show the volume in units of the compute die area of $826$mm${}^2$. We also show the modeled performance for a computer with $\pi, \beta, s$ scaled with $10^9$ compared to the A100 (the dash-dotted lines) where the same markers correspond to CG, MxM, and FFT.}
    \label{fig:a100}
\end{figure}

Using these parameters, we obtain how an \emph{Nvidia A100 homogeneous computer}, where the compute dies are stacked next to each other, would be able to perform. We show its performance in Figure~\ref{fig:a100}, assuming that the information can travel at a speed of light. As can be seen, is that in this case we are not too far away in terms of raw execution time at the scaling limit compared to a future architecture with $10^9$ higher performance than the A100. While the speed of light is a universal constant, even if we were able to communicate at this speed the fundamental limits of performance limit us from achieving much more than a 1000-fold speedup for these conventional algorithms. As such, either newer, unconventional, architectures whose computational capability can outperform the compute density of the A100 by more than $10^9$, or the algorithms we consider must change dramatically to be able to achieve more than a 1000-fold speedup at the scaling limit. This observation also holds as $n$ is increased and the only computation able to improve, both with regards to run time and scaling is MxM. 

\begin{table}[t]
\caption{Range of computer parameters for the homogeneous computer model for Frontier, Fugaku, and the Nvidia DGX GH200 with 256 superchips from Refs.~\cite{dongarra2022report,dongarra2020report,GH200,superchip}. Values marked with ${}^{\dagger}$ are experimental values.}\label{tab:computers}
\begin{center}
\begin{tabular}{llll}
\toprule
Parameter &  Frontier & Fugaku &  DGX GH200\\\midrule
$\Pi$ [Pflop/s] & $1102^\dagger$ &  $488^\dagger$ & $25.9$\\
$B$ [PB/s] &$122.3^*$ & 163 & $1.15^*$\\
$c$ [m/s]& $\approx10^{6}$ &  $\approx 10^{6}$  &  $\approx 10^{6}$  \\
$S$ [TB] & $3.1$ & 5.6 & $0.043$ \\
$D(v)$ [m] & $\sqrt{v}$ &$\sqrt{v}$ &$\sqrt{v}$\\
$V$ [m${}^2$]& 370 &1920 & 6.9 \\ \bottomrule
\end{tabular}
\end{center}
\end{table}
\begin{figure*}[t]
    \centering
    \subfloat[MxM\label{1a}]{%
       \includegraphics[width=0.333\textwidth]{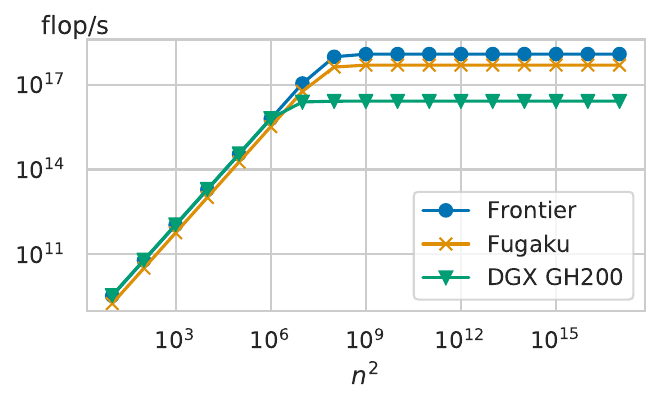}}
  \subfloat[CG\label{1b}]{%
        \includegraphics[width=0.333\textwidth]{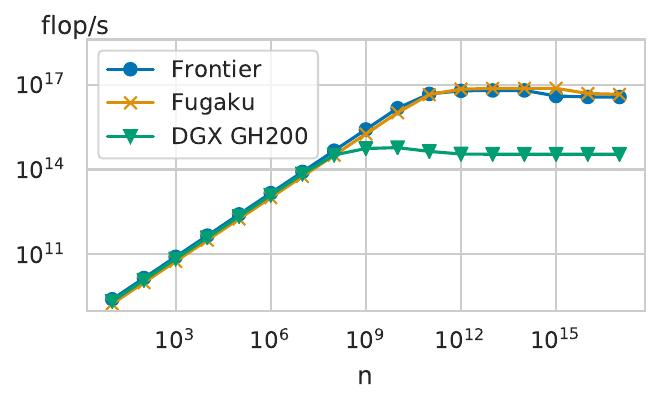}}
  \subfloat[FFT\label{1c}]{%
        \includegraphics[width=0.333\textwidth]{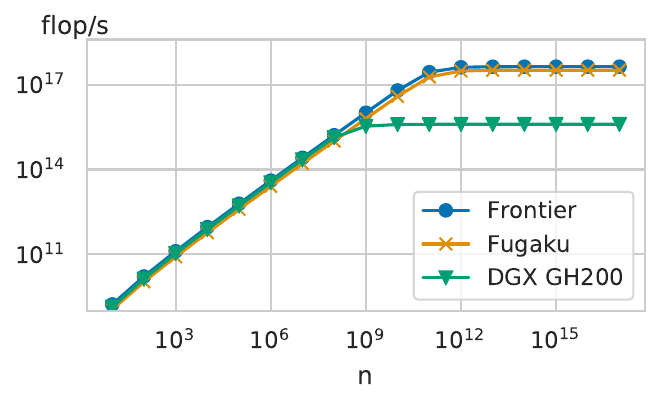}}
    \caption{Performance for the homogenous Frontier, Fugaku, and Nvidia DGX GH200 supercomputers in flop/s for MxM, CG, and FFT as $n$ increases. }
    \label{fig:front_fug_dgx}
\end{figure*}
However, current computer systems incorporating modern architectures are significantly sparser than this homogeneous computer as it is impossible to stack chips side to side. Cooling, interconnect and many other factors limit the density of the computational power. To this end, we also evaluate some current systems and their parameters within the homogeneous computer model and list them in Table \ref{tab:computers}. We model the current two largest supercomputers in the world, Frontier and Fugaku, as well as the recent Nvidia DGX GH200 AI Supercomputer, based on the new Grace Hopper Superchip. Recent interconnects such as the Slingshot 11~\cite{de2020depth} have a measured latency on the order of $\alpha =1~\mu$s, and we estimate the communication speed in these systems as $1/\alpha=10^6$ m/s.

We estimate the performance of the three computer setups in Figure~\ref{fig:front_fug_dgx} for MxM, CG, and FFT with the costs in Table~\ref{tab:costs}. What can be noted is that the behavior of Frontier and the GH200 is notably similar, with a slight edge for the GH200, for smaller workloads as they have a similar computing density with regard to $\pi=\Pi/V$ whereas Fugaku is significantly sparser. Fugaku only outperforms the two GPU-based supercomputers in the case of large CG computations, for which the larger bandwidth and cache size play an important role. This also aligns with experimental results from the HPCG benchmark~\cite{dongarra2016high} in which Fugaku has a performance of 16 Pflop/s compared to Frontiers 14 Pflop/s. For our idealized CG cost the peak performance is 73 Pflop/s and 62 Pflop/s respectively, yielding a relative performance difference of 18\%, similar to the relative difference in the actual HPCG benchmark of 14\%. Using a cost that more closely corresponds with HPCG, its preconditioner, and linear system such as the one presented in~\cite{marjanovic2015performance} we expect the absolute modeled values to be closer as well. Comparing these computers to the idealized A100 homogeneous computer it is also evident that communication, in this significantly denser setting, is the main limiting factor as the network latency of current supercomputers is already being executed on the order of $\mu$s. As such, even if computer systems compute density increases significantly in the future, the speed of light will still limit the performance of conventional large-scale workloads~\cite{markov2014limits}.

The largest difficulty in fitting model parameters for the homogeneous computer of actual systems is the correct volume and distance metric and corresponding communication speed. In this case, we focus on the area taken up by the systems, but alternatives would be to instead consider the network topology and distribution of the nodes rather than the spatial extent, and incorporate this into $D(v)$. A distance metric based on the network topology would correspond more closely to the actual limitations of sending messages across the computer. The volume metric would then rather be the number of nodes or switches used, and the distance function would need to be fitted to when one works across different switches as well as on nodes within one rack.
\begin{observation}
    The homogeneous computer model shows how we are quickly approaching the fundamental limits of computation with regard to communication speed. Even computer systems with orders of magnitude higher computing density than today will rapidly become limited by the speed of light for algorithms such as CG and FFT.
\end{observation}

\section{Discussion and Conclusion}\label{sec:conc}
In this work, we introduced the homogeneous computer model that provides us with an easily understandable tool to reason and explain general phenomena within parallel computing. It captures the scaling behavior and the considerations that must be addressed for different algorithms and shows where the main computational bottlenecks might be, effectively helping with the intuition behind high-performance algorithms and scaling behaviors. It generalizes several aspects of parallel computing, combining ideas of previous discrete computer models, the roofline model, PEM, and the cost of algorithms, and illustrates how Amdahl's and Gustafson's laws can be derived through simple physical observations about propagation speed and size of the computer system and perhaps should be reconsidered as special cases that depend on the relation between the algorithm's computational cost and computers parameters rather than general laws for parallel computing. It also illustrates both weak and strong scaling, and how the parallel efficiency for weak scaling can be manipulated by choosing the cost to $K$ to hold constant. It provides insight into how the computational cost might decrease with more processors, how the communication is alleviated when the computer is small, and how using a subset of the available resources gives the best performance for smaller problem sizes. We found that solving the minimization problem in~\eqref{eq:def} gives insight into the best possible performance for a given computer and can help algorithm and computer system design. This connection between different models, algorithmic costs, and the physical extent of the computer is the main contribution of this work. It enables an organized way of discussing scaling and performance limitations of algorithms and computer architectures and isolates fundamental limitations to parallel computing, rather than obstacles due to system design. Extending and using this work is possible in several directions, in part to accommodate more realistic values on algorithmic cost, distance functions, and computing densities, but also in its current form to assess the fundamental limits for algorithms and computer systems when projecting and predicting their future performance.

In this work, with our assessment of the model, we observed that even if the performance of future supercomputers would approach $10^9$ higher computing density than today, these computations would soon become limited by the communication speed due to the shape and physical distance within the computer. Communicating over the distances prevalent in our computer systems and networks, even at the speed of light, is quickly becoming the most prevalent performance limit even if we would overcome today's limits for transistor manufacturing. Overall, the homogeneous computer gives further indication that reconsidering classical computer systems and algorithms in pursuit of other approaches, such as approximate or randomized algorithms~\cite{mittal2016survey}, which avoid communication as far as possible, is necessary in order to perform computations that are beyond several orders of magnitude of today, even if we reach and construct an idealized homogeneous computer.
\bibliographystyle{ieeetr}
\bibliography{refs}

\begin{thebibliography}{10}

\bibitem{karp1988survey}
R.~M. Karp, {\em A survey of parallel algorithms for shared-memory machines}.
\newblock University of California at Berkeley, 1988.

\bibitem{ang2014abstract}
J.~A. Ang, R.~F. Barrett, R.~E. Benner, D.~Burke, C.~Chan, J.~Cook,
  D.~Donofrio, S.~D. Hammond, K.~S. Hemmert, S.~Kelly, {\em et~al.}, ``Abstract
  machine models and proxy architectures for exascale computing,'' in {\em 2014
  Hardware-Software Co-Design for High Performance Computing}, pp.~25--32,
  IEEE, 2014.

\bibitem{dongarra2011international}
J.~Dongarra, P.~Beckman, T.~Moore, P.~Aerts, G.~Aloisio, J.-C. Andre,
  D.~Barkai, J.-Y. Berthou, T.~Boku, B.~Braunschweig, {\em et~al.}, ``The
  international exascale software project roadmap,'' {\em The international
  journal of high performance computing applications}, vol.~25, no.~1,
  pp.~3--60, 2011.

\bibitem{alexander2020exascale}
F.~Alexander, A.~Almgren, J.~Bell, A.~Bhattacharjee, J.~Chen, P.~Colella,
  D.~Daniel, J.~DeSlippe, L.~Diachin, E.~Draeger, {\em et~al.}, ``Exascale
  applications: skin in the game,'' {\em Philosophical Transactions of the
  Royal Society A}, vol.~378, no.~2166, p.~20190056, 2020.

\bibitem{williams2009roofline}
S.~Williams, A.~Waterman, and D.~Patterson, ``Roofline: an insightful visual
  performance model for multicore architectures,'' {\em Communications of the
  ACM}, vol.~52, no.~4, pp.~65--76, 2009.

\bibitem{markov2014limits}
I.~L. Markov, ``Limits on fundamental limits to computation,'' {\em Nature},
  vol.~512, no.~7513, pp.~147--154, 2014.

\bibitem{amdahl1967validity}
G.~M. Amdahl, ``Validity of the single processor approach to achieving large
  scale computing capabilities,'' in {\em Proceedings of the April 18-20, 1967,
  spring joint computer conference}, pp.~483--485, 1967.

\bibitem{gustafson1988reevaluating}
J.~L. Gustafson, ``Reevaluating amdahl's law,'' {\em Communications of the
  ACM}, vol.~31, no.~5, pp.~532--533, 1988.

\bibitem{jia1981complexity}
H.~Jia-Wei and H.-T. Kung, ``I/o complexity: The red-blue pebble game,'' in
  {\em Proceedings of the thirteenth annual ACM symposium on Theory of
  computing}, pp.~326--333, 1981.

\bibitem{ballard2014communication}
G.~Ballard, E.~Carson, J.~Demmel, M.~Hoemmen, N.~Knight, and O.~Schwartz,
  ``Communication lower bounds and optimal algorithms for numerical linear
  algebra,'' {\em Acta Numerica}, vol.~23, pp.~1--155, 2014.

\bibitem{arge2008fundamental}
L.~Arge, M.~T. Goodrich, M.~Nelson, and N.~Sitchinava, ``Fundamental parallel
  algorithms for private-cache chip multiprocessors,'' in {\em Proceedings of
  the twentieth annual symposium on Parallelism in algorithms and
  architectures}, pp.~197--206, 2008.

\bibitem{demaine2018red}
E.~D. Demaine and Q.~C. Liu, ``Red-blue pebble game: Complexity of computing
  the trade-off between cache size and memory transfers,'' in {\em Proceedings
  of the 30th on Symposium on Parallelism in Algorithms and Architectures},
  pp.~195--204, 2018.

\bibitem{kwasniewski2019red}
G.~Kwasniewski, M.~Kabi{\'c}, M.~Besta, J.~VandeVondele, R.~Solc{\`a}, and
  T.~Hoefler, ``Red-blue pebbling revisited: near optimal parallel
  matrix-matrix multiplication,'' in {\em Proceedings of the International
  Conference for High Performance Computing, Networking, Storage and Analysis},
  pp.~1--22, 2019.

\bibitem{smith2017tight}
T.~M. Smith, B.~Lowery, J.~Langou, and R.~A. van~de Geijn, ``A tight i/o lower
  bound for matrix multiplication,'' {\em arXiv preprint arXiv:1702.02017},
  2017.

\bibitem{rico2019survey}
J.~A. Rico-Gallego, J.~C. D{\'\i}az-Mart{\'\i}n, R.~R. Manumachu, and A.~L.
  Lastovetsky, ``A survey of communication performance models for
  high-performance computing,'' {\em ACM Computing Surveys (CSUR)}, vol.~51,
  no.~6, pp.~1--36, 2019.

\bibitem{culler1993logp}
D.~Culler, R.~Karp, D.~Patterson, A.~Sahay, K.~E. Schauser, E.~Santos,
  R.~Subramonian, and T.~Von~Eicken, ``Logp: Towards a realistic model of
  parallel computation,'' in {\em Proceedings of the fourth ACM SIGPLAN
  symposium on Principles and practice of parallel programming}, pp.~1--12,
  1993.

\bibitem{hoefler2010loggopsim}
T.~Hoefler, T.~Schneider, and A.~Lumsdaine, ``Loggopsim: simulating large-scale
  applications in the loggops model,'' in {\em Proceedings of the 19th ACM
  International Symposium on High Performance Distributed Computing},
  pp.~597--604, 2010.

\bibitem{fisher1988your}
D.~C. Fisher, ``Your favorite parallel algorithms might not be as fast as you
  think,'' {\em IEEE transactions on computers}, vol.~37, no.~02, pp.~211--213,
  1988.

\bibitem{dufek2021extended}
A.~S. Dufek, J.~R. Deslippe, P.~T. Lin, C.~J. Yang, B.~G. Cook, and J.~Madsen,
  ``An extended roofline performance model with pci-e and network ceilings,''
  in {\em 2021 International Workshop on Performance Modeling, Benchmarking and
  Simulation of High Performance Computer Systems (PMBS)}, pp.~30--39, IEEE,
  2021.

\bibitem{cardwell2019extended}
D.~Cardwell and F.~Song, ``An extended roofline model with
  communication-awareness for distributed-memory hpc systems,'' in {\em
  Proceedings of the International Conference on High Performance Computing in
  Asia-Pacific Region}, pp.~26--35, 2019.

\bibitem{abdelfattah2021survey}
A.~Abdelfattah, H.~Anzt, E.~G. Boman, E.~Carson, T.~Cojean, J.~Dongarra,
  A.~Fox, M.~Gates, N.~J. Higham, X.~S. Li, {\em et~al.}, ``A survey of
  numerical linear algebra methods utilizing mixed-precision arithmetic,'' {\em
  The International Journal of High Performance Computing Applications},
  vol.~35, no.~4, pp.~344--369, 2021.

\bibitem{solomonik2017trade}
E.~Solomonik, E.~Carson, N.~Knight, and J.~Demmel, ``Trade-offs between
  synchronization, communication, and computation in parallel linear algebra
  computations,'' {\em ACM Transactions on Parallel Computing (TOPC)}, vol.~3,
  no.~1, pp.~1--47, 2017.

\bibitem{hoemmen2010communication}
M.~Hoemmen, {\em Communication-avoiding Krylov subspace methods}.
\newblock University of California, Berkeley, 2010.

\bibitem{abelson2000amorphous}
H.~Abelson, D.~Allen, D.~Coore, C.~Hanson, G.~Homsy, T.~F. Knight~Jr,
  R.~Nagpal, E.~Rauch, G.~J. Sussman, and R.~Weiss, ``Amorphous computing,''
  {\em Communications of the ACM}, vol.~43, no.~5, pp.~74--82, 2000.

\bibitem{nagpal2004engineering}
R.~Nagpal and M.~Mamei, ``Engineering amorphous computing systems,'' in {\em
  Methodologies and software engineering for agent systems: The agent-oriented
  software engineering handbook}, pp.~303--320, Springer, 2004.

\bibitem{beal2004programming}
J.~Beal, ``Programming an amorphous computational medium,'' in {\em
  International Workshop on Unconventional Programming Paradigms},
  pp.~121--136, Springer, 2004.

\bibitem{lloyd2000ultimate}
S.~Lloyd, ``Ultimate physical limits to computation,'' {\em Nature}, vol.~406,
  no.~6799, pp.~1047--1054, 2000.

\bibitem{karp2022reducing}
M.~Karp, N.~Jansson, A.~Podobas, P.~Schlatter, and S.~Markidis, ``Reducing
  communication in the conjugate gradient method: a case study on high-order
  finite elements,'' in {\em Proceedings of the Platform for Advanced
  Scientific Computing Conference}, pp.~1--11, 2022.

\bibitem{takahashi2019fast}
D.~Takahashi, {\em Fast Fourier transform algorithms for parallel computers}.
\newblock Springer, 2019.

\bibitem{dongarra1997top500}
J.~J. Dongarra, H.~W. Meuer, E.~Strohmaier, {\em et~al.}, ``Top500
  supercomputer sites,'' {\em Supercomputer}, vol.~13, pp.~89--111, 1997.

\bibitem{ampere}
``Nvidia a100 tensor core gpu architecture,'' white paper, Nvidia, Santa Clara,
  United States.

\bibitem{dongarra2022report}
J.~Dongarra and A.~Geist, ``Report on the oak ridge national laboratory’s
  frontier system,'' {\em Univ. of Tennessee, Knoxville, Tech. Rep. Tech Report
  No. ICL-UT-22-05}, 2022.

\bibitem{dongarra2020report}
J.~Dongarra, ``Report on the fujitsu fugaku system,'' {\em University of
  Tennessee-Knoxville Innovative Computing Laboratory, Tech. Rep. ICLUT-20-06},
  2020.

\bibitem{GH200}
``{NVIDIA DGX GH200 AI Supercomputer},'' white paper, Nvidia, Santa Clara,
  United States, June 2023.
\newblock Accessed: September, 2023.

\bibitem{superchip}
``{NVIDIA GH200 Grace Hopper Superchip Datasheet},'' white paper, Nvidia, Santa
  Clara, United States, August 2023.
\newblock Accessed: September, 2023.

\bibitem{de2020depth}
D.~De~Sensi, S.~Di~Girolamo, K.~H. McMahon, D.~Roweth, and T.~Hoefler, ``An
  in-depth analysis of the slingshot interconnect,'' in {\em SC20:
  International Conference for High Performance Computing, Networking, Storage
  and Analysis}, pp.~1--14, IEEE, 2020.

\bibitem{dongarra2016high}
J.~Dongarra, M.~A. Heroux, and P.~Luszczek, ``High-performance
  conjugate-gradient benchmark: A new metric for ranking high-performance
  computing systems,'' {\em The International Journal of High Performance
  Computing Applications}, vol.~30, no.~1, pp.~3--10, 2016.

\bibitem{marjanovic2015performance}
V.~Marjanovi{\'c}, J.~Gracia, and C.~W. Glass, ``Performance modeling of the
  hpcg benchmark,'' in {\em High Performance Computing Systems. Performance
  Modeling, Benchmarking, and Simulation: 5th International Workshop, PMBS
  2014, New Orleans, LA, USA, November 16, 2014. Revised Selected Papers 5},
  pp.~172--192, Springer, 2015.

\bibitem{mittal2016survey}
S.~Mittal, ``A survey of techniques for approximate computing,'' {\em ACM
  Computing Surveys (CSUR)}, vol.~48, no.~4, pp.~1--33, 2016.

\end{thebibliography}
\end{document}